\documentclass[arxiv,superscriptaddress,twocolumn]{revtex4-1}

\usepackage{lineno,hyperref}

\usepackage{graphicx,ulem}
\usepackage{bmpsize}
\usepackage{amsfonts}
\usepackage[usenames, dvipsnames]{color}

\newcommand{\ben}{\begin{eqnarray}}
\newcommand{\een}{\end{eqnarray}}

\newcommand{\bef}{\begin{figure}[h!bt]\centering}
\newcommand{\eef}{\end{figure}}
\newcommand{\bet}{\begin{table}[hbt]\centering}
\newcommand{\eet}{\end{table}}

\begin{document}

\title{Charge ordering and ferrimagnetism in the strongly correlated $\beta$-V$_2$PO$_5$ single crystal}

\author{Jie Xing}
\affiliation{Department of Physics and Astronomy and California NanoSystems Institute, University of California, Los Angeles, CA 90095, USA}
\author{Huibo Cao}
\affiliation{Neutron Scattering Division, Oak Ridge National Laboratory, Oak Ridge, TN 37831, USA}
\author{Arpita Paul}
\affiliation{Department of chemical engineering and materials science,University of Minnesota, MN 55455, USA}
\author{Chaowei Hu}
\affiliation{Department of Physics and Astronomy, University of California, Los Angeles, CA 90095, USA}
\author{Hsin-Hua Wang}
\affiliation{Department of Physics and Astronomy, University of California, Los Angeles, CA 90095, USA}
\author{Yongkang Luo}
\affiliation{Department of Physics and Astronomy, University of California, Los Angeles, CA 90095, USA}
\author{Raj Chaklashiya}
\affiliation{Department of Physics and Astronomy, University of California, Los Angeles, CA 90095, USA}
\author{Jared M. Allred}
\affiliation{Department of Chemistry and Biochemistry, University of Alabama, Tuscaloosa, AL 35487, USA}
\author{Stuart Brown}
\affiliation{Department of Physics and Astronomy, University of California, Los Angeles, CA 90095, USA}
\author{Turan Birol}
\affiliation{Department of chemical engineering and materials science,University of Minnesota, MN 55455, USA}
\author{Ni Ni}
\email{Corresponding author: nini@physics.ucla.edu}
\affiliation {Department of Physics and Astronomy and California NanoSystems Institute, University of California, Los Angeles, CA 90095, USA}

\begin{abstract}
A combined study of transport, thermodynamic, neutron diffraction, nuclear magnetic resonance measurements and first principles calculation were performed for $\beta$-V$_2$PO$_5$ single crystal. It was shown to be a semiconductor with a band gap of 0.48 eV, undergoing a charge ordering (unusual V$^{2+}$ and V$^{3+}$) phase transition accompanied by a tetragonal to monoclinic structural distortion at 610 K and a paramagnetic to ferrimagnetic phase transition at 128 K with a propagation vector of $\textbf{k} = 0$. The easy axis is in the monoclinic $ac$ plane pointing 47(9)$^\circ$ away from the monoclinic $a$ axis. This collinear ferrimagnetic structure and anisotropic isothermal magnetization measurements suggest weak magnetic anisotropy in this compound. The first principles calculations indicate that the intra-chain interactions in the face-sharing VO$_6$ chains dominate the magnetic hamiltonian and identify the $\Gamma_5^+$ normal mode of the lattice vibration to be responsible for the charge ordering and thus the structural phase transition.

\end{abstract}


\maketitle

\section{Introduction}
Charge ordering(CO), the long-range ordering of transition metal ions with different oxidization states, is a prominent feature in mixed valent 3$d$ transition metal oxides \cite{COTO}. Due to the strong Coulomb interaction in the charge ordering state, a high symmetry to low symmetry structural distortion can occur, accompanied with the sudden enhancement in the electrical resistivity arising from the charge localization. The competition between this charge disproportionation and the exchange interactions among magnetic transition metal ions has led to emergent phenomena, such as colossal magnetoresistance in RE$_{1-x}$A$_x$MnO$_3$ (RE = rare earth, A = alkaline earth) \cite{MnO1,MnO2}, superconductivity in $\beta$-Ag$_{0.33}$V$_2$O$_5$\cite{vosc}, etc.

The vanadium phosphorus oxide system (V-P-O) has distinct structural stacking and variable valences of vanadium, providing a great avenue to investigate the structure-property relationship, enriching our understanding on the competition of CO and various exchange interaction. The fundamental building blocks of the V-P-O system consist of VO$_6$ octahedra or VO$_4$ tetrahedra linked by PO$_4$ tetrahedra with valence P$^{5+}$. Rich 3$d$ vanadium magnetism and valences have been observed. For example, VPO$_4$ with V$^{3+}$ ions, containing one dimensional chains of edge-sharing VO$_6$ octahedra, undergoes an incommensurate antiferromagnetic (AFM) phase transition at 26 K and then a commensurate AFM phase transition at 10.3 K \cite{vpo4}. $\alpha$-VO(PO$_3$)$_2$ with one dimensional chains of corner-sharing VO$_6$ octahedra, is AFM at 1.9 K with valence V$^{4+}$ \cite{VO(PO3)2nmr}. Mixed valence V$^{3+}$ and V$^{4+}$ antiferromagnetically couple together below 5 K in V$_2$(VO)(P$_2$O$_7$)$_2$, where segments of edge-sharing VO$_4$ tetrahedra and VO$_6$ octahedra exist \cite{v2vop2o72}. Alternating V$^{4+}$ spin-chain model can be used to describe the magnetism in (VO)$_2$P$_2$O$_7$ with corner and edge-sharing VO$_6$ octahedra ladders \cite{(VO)2P2O7 theory1,(VO)2P2O7 theory2, (VO)2P2O7 neotron,(VO)2P2O7nmr,(VO)2P2O7nmr2}.

In this article, we investigated $\beta$-V$_2$PO$_5$. In the tetragonal phase in Ref. \cite{v2o5p structure1} (Fig. 3(b)), it contains chains of face-sharing VO$_6$ octahedra linked by PO$_4$ tetrahedra. These chains are stacked in layers along the $c$ axis, running alternately along the $a$ or $b$ axis in the adjacent layers (Fig. 3b).
This material is intriguing in three aspects. Firstly, the valence analysis with P$^{5+}$ and O$^{2-}$ indicates remarkably low valence V$^{2.5+}$ in this compound, which may suggest possible CO of V$^{3+}$ and very uncommon valence V$^{2+}$ \cite{johnston}. Secondly, the face-sharing VO$_6$ octahedra in the building block is very rare for vanadium oxides, implying unusually strong intra-chain interaction between V ions. What's more, although the parallel chains in each layer do not share any oxygen, the perpendicularly-running chains in the neighboring layers are corner-sharing. As a result, the other important magnetic interaction is the inter-chain interaction between the corner-sharing V ions in neighboring layers. Thirdly, a recent first-principles calculation suggested that the $\beta$-V$_2$PO$_5$ is a ferromagnetic (FM) topological Weyl and node-line semimetal without any trivial band at the Fermi level \cite{v2o5p}, being a great material platform to study the emergent phenomena in FM topological semimetals.

Despite of these remarkable aspects we discussed above, neither physical properties nor possible structural distortion has been investigated for $\beta$-V$_2$PO$_5$, therefore, we performed a combined study of the single crystalline $\beta$-V$_2$PO$_5$ by x-ray and neutron diffraction as well as NMR, transport and thermodynamic measurements. We discovered that $\beta$-V$_2$PO$_5$ is a semiconductor with a band gap of 0.48 eV. Upon cooling, a charge ordering phase transition accompanied with a tetragonal to monoclinic structural phase transition occurs at 610 K, followed by a long range ferrimagnetic phase transition below 128 K.
\section{Experimental methods}

Precursor $\beta$-V$_2$PO$_5$ powder was made by solid state reaction. V$_2$O$_5$ powder and phosphorus chunks were weighed according to the stoichiometric ratio 1 : 1 and sealed in a quartz tube under vacuum. The ampule was slowly heated up to 600$^\circ $C and dwelled for 2 hours, and then was increased to 1000$^\circ $C and stayed for 2 days before it was quenched in water. The resultant $\beta$-V$_2$PO$_5$ ($\sim$ 2 g) powder and iodine flakes (10 mg / cm$^3$) were loaded into a 15-cm long quartz tube and sealed under vacuum. Single crystals of $\beta$-V$_2$PO$_5$ were then grown by chemical vapor transport method \cite{v2o5p structure1}. The hot end was set at 1000$^\circ $C and the cold end was set at 900$^\circ $C. After two weeks, quite a few sizable three dimensional single crystals ($\sim$4 mm$\times$4mm$\times$2mm) were found at the cold end.
The inset of Fig. 1(a) shows a $\beta$-V$_2$PO$_5$ single crystal against 1 mm scale.

Throughout the paper, $ab$ plane is the plane where chains locate in. The (hkl)$_T$ means the peak indexed in the tetragonal structure while (hkl)$_M$ means the peak indexed in the monoclinic structure.

Magnetic properties were measured in a Quantum Design (QD) Magnetic Properties Measurement System (MPMS3). A single crystal around 20 mg with a polished $ab$ surface and a single crystal with as-grown (011) surface were used. Temperature dependent heat capacity was measured in a QD Dynocool Physical Properties Measurement System (Dynoccol PPMS) using the relaxation technique at zero field.
To enhance the thermal contact and lower the measurement time, the $\beta$-V$_2$PO$_5$ single crystal was ground into powder and then mixed with silver powder according to the mass ratio of 1 : 1 . The heat capacity of $\beta$-V$_2$PO$_5$ was then obtained by subtracting the heat capacity of silver \cite{silver}.
Below 200 K, the two wire ETO method was used for the electric resistivity measurement in PPMS. From 200 K to 400 K, the electric resistivity was measured with standard four-point method while above 400 K, it was measured in a homemade high temperature resistivity probe.

Single crystal neutron diffraction was performed at the HB-3A four-circle diffractometer equipped with a 2D detector at the High Flux Isotope Reactor(HFIR) at ORNL. Neutron wavelength of 1.546~\AA~was used from a bent perfect Si-220 monochromator \cite{hb3a}.  The pyrolytic graphite (PG) filter was used before the sample to reduce the half-$\lambda$ neutrons. Representational analysis with \textit{SARAh} \cite{sarah} was run to search for the possible magnetic symmetries. The nuclear and magnetic structure refinements were carried out with the FullProf Suite\cite{fullprof}.
 Powder X-ray diffraction measurements were performed using a PANalytical Empyrean diffractometer (Cu K$\alpha$ radiation). Using the Fullprof suit\cite{fullprof}, Rietveld refinement was carried out to refine the powder X-ray diffraction data with the crystal structure determined by single crystal neutron diffraction.

Nuclear magnetic resonance (NMR) measurement was done under a fixed magnetic field of approximately 8.5 T, applied along the direction perpendicular to the $ab$ plane, where the chains locate in. The spectra were collected by performing an optimized $\pi/2$-$\tau$-$\pi$ spin-echo pulse sequence. The spin-lattice relaxation time $T_1$ was measured by integration of the phase corrected real part of the spin echo using the saturation-recovery technique cite and spin echo decay time $T_2$ was measured by altering $\tau$ in the sequence.
Spin-lattice relaxation time $T_1$ is obtained by the magnetization recovery fitting to a single exponential form.

First principles Density Functional Theory calculations were performed to compare the energies of different magnetic configurations. We used PAW as implemented in VASP with PBEsol exchange correlation functional \cite{VASP1, VASP2, PBEsol}. A $8\times 8\times 8$ k-point grid and energy cut off of 500 eV ensures convergence in the primitive cell with 4 formula units.

\section{Experimental results}
\subsection{Magnetic, transport and thermodynamic properties}

\begin{figure*}
\includegraphics[width=7in]{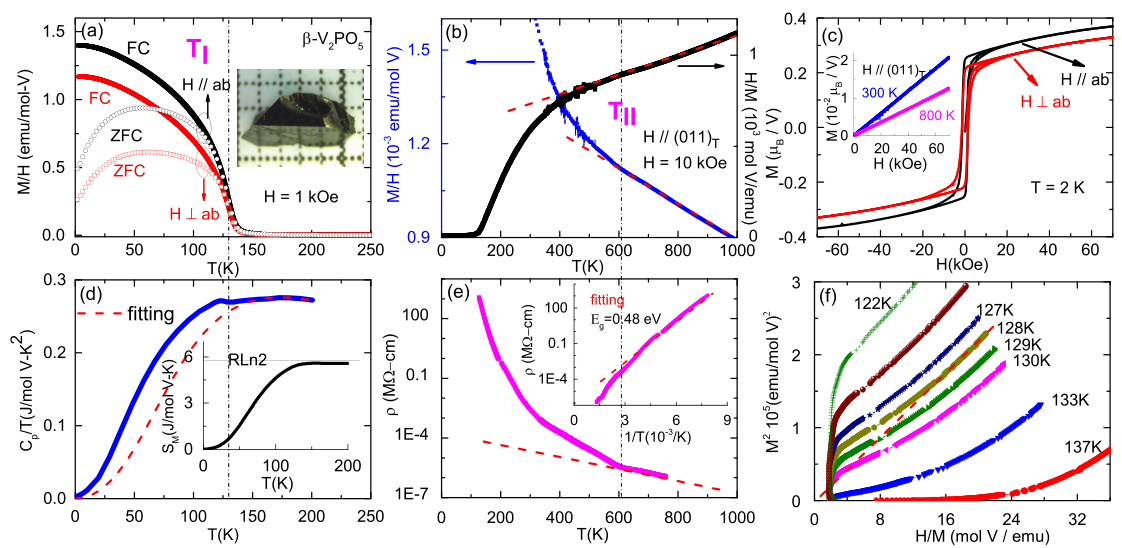}
\caption {(a) ZFC and FC $M/H$ vs. $T$ under $H$ = 1 kOe with $H // ab$ and $H \perp ab$ from 2 K to 250 K. Inset: picture of $\beta$-V$_2$O$_5$P single crystal against 1-mm scale. (b) $H/M$ and $M/H$ vs. $T$ under $H$ =10 kOe with $H // (0 1 1)_T$ 2 K to 1000 K. The red line is the Curie-Weiss fit. (c) Isothermal $M(H)$ curves at 2 K with $H // ab$ and $H \perp ab$. Inset: $M(H)$ curves at 300 K and 800 K along with $H // (0 1 1)_T$. (d) Specific heat $C_p$ vs. $T$ from 2 K to 200 K. The red line is the fitting curve by Debye model. Inset: Temperature dependence of magnetic entropy.  (e) Resistivity $\rho$ vs. $T$ from 130 K to 760 K. The red line emphasizes the transition at $T_{II}$. Inset: $\rho$ vs. $1/T$. Red line: the fitting curve using the thermal excitation model. (f) The Arrott plot at various temperatures from 122 K to 137 K.}
\label{fig2}
\end{figure*}

Figure 1 (a)-(c) show the anisotropic magnetic properties of $\beta$-V$_2$PO$_5$.
Figure~1(a) presents the temperature dependent $M/H$ taken at $H$ = 1 kOe from 2 K to 250 K in zero-field-cooled (ZFC) warming and field-cooled mode with $H$ parallel and perpendicular to the $ab$ plane. The sharp upturns of the curves and the bifurcation in ZFC and FC data for both directions below $T_I$ indicate the existence of ferromagnetic component. The smooth ZFC and FC curves suggest no other magnetic transition below $T_I$. Comparing with the other V-O-P materials, the magnetic transition temperature is quite high \cite{vpo4,VO(PO3)2nmr,v2vop2o72}, suggesting strong exchange interactions.
Figure 1. (b) shows the temperature dependent $M/H$ (blue) and $H/M$ (black) measured from 300 K to 1000 K with $H$ // (0 1 1)$_T$ at $H$ = 10 kOe. Firstly, we see a subtle but discernable enhancement in $M/H$ at the characteristic temperature $T_{II}$ = 610 K, suggesting a possible phase transition here. Secondly, upon cooling, linear Curie-Weiss behavior can be clearly seen from 1000 K to 500 K in $H/M$. By fitting $H/M$ from 1000 K to 500 K using the Curie-Weiss formula $H/M=C/(T-\theta_{cw})$, where $C$ is the Curie constant and $\theta_{cw}$ is the Weiss temperature, we obtained $\mu_{eff}$ = 3.7(2) $\mu_{B}$/V and $\theta_{cw}=-900$ K. The $\mu_{eff}$ is larger than the one of V$^{2.5+}$ but comparable to the one of $V^{2+}$. The large negative $\theta_{cw}$ with $|\theta_{cw}/T_I| \sim$ 7.2, suggests strong antiferromagnetic interaction. Thirdly, the $H/M$ from $T_I$ to 500 K shows a crossover concave behavior with temperature.

Figure 1 (c) shows the anisotropic field dependent magnetization $M(H)$ taken at 2 K with $H // ab$ and $H \perp ab$. The crystal orientation was determined by x-ray diffraction (Fig. S1) \cite{supp}. Clear hysteresis can be observed in both directions, confirming the existence of ferromagnetic component. Both curves show very similar shape and magnitude, suggesting weak magnetic anisotropy in this system. The coercive fields are around 1.3 kOe for both. The remanent moments are 0.22 $\mu_B$/V for $H \perp ab$ and 0.25 $\mu_B$/V for $H // ab$ and the saturation moments are 0.27 $\mu_B$/V for $H \perp ab$ and 0.31 $\mu_B$/V for $H // ab$, which are so much smaller than the saturation moment of V$^{2+}$ ($d^3$) and V$^{3+}$ ($d^2$) ions, suggesting that instead of ferromagnetism, this material is likely ferrimagnetic or canted antiferromagnetic below $T_I$. The inset of Fig. 1(c) shows the $M(H)$ curves taken at 300 K and 800 K with
$H$ // (0 1 1)$_T$. Both curves are linear with the applied magnetic field without hysteresis. The Arrott plot ($M^2$ vs. $H/M$) has been widely used to determine the ferromagnetic phase transition temperature \cite{arrot1, arrot2}, where the curve of $M^2$ vs. $H/M$ passes through
the origin of the plot at the transition temperature. To determine the value of $T_I$, isothermal $M(H)$ curves are measured from 122 K to 137 K. The Arrott plot are calculated and shown in Fig. 1(f), which suggests that $T_I \sim 128$ K.

Figure 1 (d) shows the temperature dependent $C_p/T$ data (blue) taken from 2 K to 200 K. A heat capacity anomaly featuring a second order phase transition appears around 128 K, accompanying with the magnetic phase transition observed in Fig. 1 (a). To estimate the magnetic entropy, Debye model is used to fit the heat capacity to provide the non-magnetic background. The fitted curve is shown in red in Fig. 1(d) and the fitted Debye temperature is $\Theta_D=620$ K. By subtracting the non-magnetic background from fitting, we obtained the magnetic entropy as $S_M$ = 5.6 J/mol V-K$^2$. This value is significantly smaller than $R$Ln4 of V$^{2+}$ and $R$Ln3 of V$^{3+}$, but rather approximate $R$Ln2. This may be caused by the strong V-O covalency which lowers the moment size of V or a result of the entropy release above 128 K due to the chain structure and strong intrachain interaction \cite{entropy dimension1,entropy dimension2}.

Figure 1(e) shows the resistivity ($\rho$) of the $\beta$-V$_2$PO$_5$ single crystals vs. temperature from 130 K up to 760 K. Instead of the semi-metal suggested by the theoretical prediction \cite{v2o5p}, it is a semiconductor. A semiconductor to semiconductor phase transition is discernable at 610 K, which confirms the possible phase transition at $T_{II}$ suggested by the subtle susceptibility increase shown in Fig. 1 (b). $\rho$ vs. reciprocal temperature is plotted in the inset of Fig.~1(d). By fitting the data between 280 K to 130 K with the thermal excitation model $\rho(T)=\rho(0)$exp$(E_g/2k_BT)$, the estimated gap size of $\beta$-V$_2$PO$_5$ is 0.48 eV. The gap value is similar to 0.45-0.57 eV of the Vanadium phosphate glass \cite{VPOglass}.

\subsection{Structural phase transition and charge ordering}

Based on the transport, magnetic and heat capacity measurements, we have shown that $\beta$-V$_2$PO$_5$ has one phase transition at 610 K and the other magnetic phase transition at 128 K. To investigate the nature of these two phase transitions, single crystal neutron diffraction and NMR measurements are performed, which are summarized in Fig. 2.

Figure 2 (a) and (c) show the order parameter plot of the (1 1 4)$_T$ neutron peak up to 650 K and (1 1 0)$_T$ neutron peak up to 450 K, respectively and Fig. 2 (b) presents the rocking curve scan for the (1 1 4)$_T$ peak. It is twinned structure below 610 K, so we keep the tetragonal index for convenience.
The order parameter plot of the relative stronger peak (1 1 4)$_T$ (Fig. 2(a)) indicates two phase transitions occurring at 610 K and 128 K, respectively, which is consistent with the order parameter plot of the (1 1 0)$_T$ peak (Fig. 2(c)) and (1 0 1)$_T$ peak (Fig. S2) \cite{supp}. The full data were collected at 4.5 K, 300 K, and 650 K to cover all three phase regions. At 650 K, the data can be well fitted in $I$41/$amd$ symmetry (Table I). Since both (1 1 0)$_T$ and (1 1 4)$_T$ peaks are symmetry disallowed reflections in the tetragonal $I$41/$amd$ structure, the fact that we observed these peaks at room temperature (Fig. 2(b)) suggests possible structural/magnetic phase transitions at 610 K.

\begin{figure}
\includegraphics[width=3.5in]{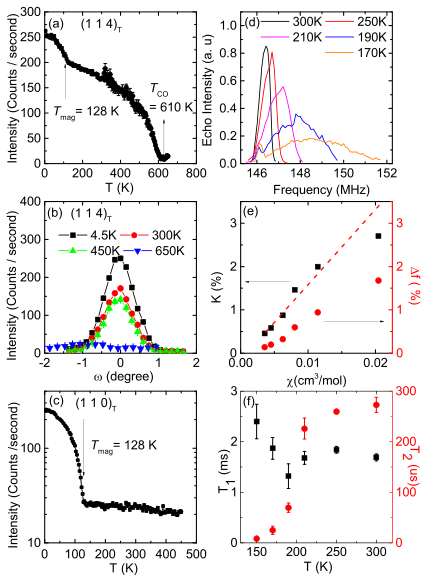}
\caption {(a) The (1 1 4)$_T$ neutron peak intensity vs. T. (b) The (1 1 4)$_T$ neutron peak intensity vs. $\omega$. (c) The (1 1 0)$_T$ neutron peak intensity vs. T. (d) P NMR frequency spectra. The spectra are conserved after corrections for T$_2$. (e) Knight shift K and the peak width $\Delta$F obtained from (a) vs. magnetic susceptibility $\chi$. Dashed line shows a linear fittings of K. (f) Spin-lattice relaxation time $T_1$ and spin-spin relaxation time $T_2$ vs. T.
}\label{}
\end{figure}

\begin{figure}
\includegraphics[width=3.5in]{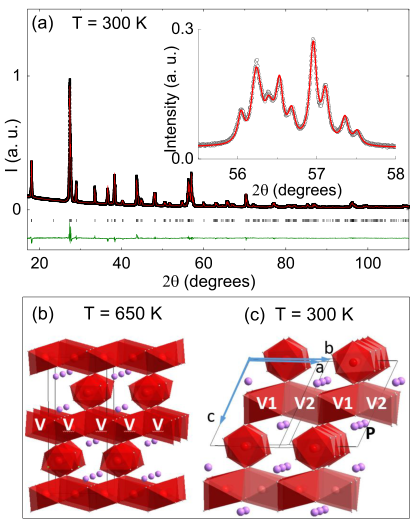}
\caption {(a) The experimental and refined powder X-ray diffraction patterns for $\beta$-V$_2$O$_5$P at 300 K. Black: experimental pattern. Red: refined pattern. green: the difference between the experimental and refined patterns. Black ticks: the Bragg peak positions in the monoclinic structure. Inset: enlarged view from 55.5$^\circ$ to 58$^\circ$. (b)(c): The crystal structure of $\beta$-V$_2$O$_5$P at 650 K (b) and 300 K (c). }\label{fig2}
\end{figure}

\begin{table}
\caption{The crystal structure of the $\beta$-V$_2$OPO$_4$ phase at 4.5 K, 300 K and 650 K, respectively. }
\begin{tabular}{p{2cm}p{2cm}p{2cm}p{2cm}}
\hline\hline
\multicolumn{4}{c}{$\beta$-V$_2$PO$_5$ at 4.5 K ~~monoclinic  \textit{C2/c}  }\\
\hline
a= 7.563{\AA}&b=7.563{\AA}&c=7.235 {\AA} &$\beta$= 121.51$^\circ$\\
R$_{F^2}$=0.0691& wR$_{F^2}$=0.0841&R$_{F}$=0.044&$\chi^2$=19.2\\
\hline
site  & x/a   & y/b & z/c \\
\hline
V1   & 0   &  1/2   &  0 \\
V2   & 1/4     & 1/4    &  0 \\
O1   & 0.066(3)   & 0.749(2)  & 0.632(2) \\
O2   & 0.317(2)     &  0.496(2)  & 0.598(1) \\
O3   & 0     &  0.652(2)   & 1/4  \\
P    & 0     &  0.121(2)  & 1/4  \\
\hline
\hline

\multicolumn{4}{c}{$\beta$-V$_2$PO$_5$ at 300 K~~ monoclinic \textit{C2/c} }\\
\hline
a= 7.570{\AA}&b=7.570{\AA}&c=7.232 {\AA}&$\beta$= 121.56$^\circ$\\
R$_{F^2}$=0.0734& wR$_{F^2}$=0.0956&R$_{F}$=0.0473&$\chi^2$=25.1\\
\hline
site  & x/a   & y/b & z/c \\
\hline
V1   & 0   &  1/2     &  0 \\
V2   & 1/4     & 1/4    &  0 \\
O1   & 0.065(3)   & 0.749(2)  & 0.630(2) \\
O2   & 0.318(2)     &  0.496(2)  & 0.599(1) \\
O3   & 0     &  0.652(2)   & 1/4  \\
P    & 0     &  0.119(3)  & 1/4  \\
\hline
\hline
\multicolumn{4}{c}{$\beta$-V$_2$PO$_5$ at 650K ~~Tetragonal \textit{I41/amd} }\\
\hline
a= 5.357(2){\AA}&b= 5.357(2){\AA}&c=12.373(4){\AA}\\
R$_{F^2}$=0.0707& wR$_{F^2}$=0.0888&R$_{F}$=0.0421&$\chi^2$=5.12\\
\hline
site  & x/a   & y/b & z/c \\
\hline
V1   & 1/4    &  1/4    &  1/4 \\
P1   & 0    & 3/4    &  1/8 \\
O1   & 0   & -0.012(1)  & 0.193(4) \\
O2   & 0     &  3/4  & 5/8 \\
\hline\hline
\end{tabular}
\label{tab.1}
\end{table}

To identify if the phase between 128 K and 610 K has a magnetic component, phosphorus-31 NMR ($^{31}P$-NMR) measurements were carried out. These are summarized in Fig. 2(e)-(f). In Fig. 2(d), we report the $^{31}$P-NMR spectra at various temperatures from 170 K to 300 K. $^{31}$P nuclear spin $I=1/2$, and all sites are equivalent for $B \perp ab$. At 300 K, the spectra shift from the Larmor frequency by $K_s=0.493\pm0.002\%$, where the mean and uncertainty were calculated using the gaussian fitting. This value is on the same order of another VPO sample with V$^{3+}$ and about twice as much as that with V$^{4+}$ \cite{NMR1,NMR2}, where a similar shift and broadening were observed. The observations are interpreted as evidence for no long range magnetic ordering in the intermediate, charge-ordered phase.
Below 190 K, a minor absorption peak at 146.4 MHz is resolved. Since it accounts for only 3\% of the total spin intensities, we expect the signal to be extrinsic due to the sites at twin boundaries and exclude it in our analysis. From  the  slope  of  K-$\chi$ plot  shown  in  Fig.2 (e), we estimate the hyperfine coupling constant to be $A_{cc}= (10.23\pm0.87)$ kOe/$\mu_B$, which is transferred from the unpaired  electrons  from  the  second  nearest  neighbors  of P \cite{NMR3}.
Figure 2(f) shows the temperature dependence of the spin-lattice relaxation time $T_1$ and spin-spin relaxation time $T_2$. The relatively short and constant $T_1$ is consistent with a paramagnetic phase from 150 k to 300 K \cite{MORIYA}. Meanwhile, $T_2$ starts dropping rapidly below 210 K as the system approaches the 128 K transition. This behavior is associated with slow longitudinal fluctuations, which are likely related with the onset of the 128 K magnetic phase transition. This is also consistent with the broadening observed in Fig. 2(d) which appears as the magnetic correlation develops upon cooling. However, since 1/$T_2$ is on the order of a few to hundreds of KHz, we conclude that most of our broadening, which is on the order of several MHz, is from the inhomogeneous internal field.

Since NMR shows that the phase transition at 610 K is not of magnetic origin, the phase transition at 610 K should be a structural distortion. By including four twinned structure domains, the room temperature neutron data can be well fitted with the $C$2/$c$ monoclinic symmetry, suggesting a tetragonal to monoclinic phase transition at 610 K. Using this crystal structure, we refined our powder X-ray diffraction taken at room temperature and obtained a very good fit as shown in Fig. 3(a). The detailed crystal structures at 4.5 K, 300 K and 650 K are summarized in Table I. The high temperature tetragonal and low temperature monoclinic structures are visualized in Fig. 3(b) and (c), respectively.
In the monoclinic phase, the monoclinic $c$ axis is 121.45$^\circ$ from the $ab$ plane which is the plane where the chains sit in. The unique V site in the tetragonal structure separates into V1 and V2 sites with the V1 atoms and V2 atoms alternately locating along each chain direction (V1 and V2 sites are labeled in Fig.~3(c)), as a result, the average bond length of VO$_6$ octahedra on V1 site increases while that on V2 site decreases. Bond-valence analysis of the monoclinic crystal structure assigns charges of 2.0 and 2.9 to $V_1$ and $V_2$ respectively, which is a smoking gun proof of the charge order \cite{bondvalence_parameters, bondvalence_parameters1}.

\subsection{Ferrimagnetic structure below 128 K}

\begin{figure}
\includegraphics[width=3.5in]{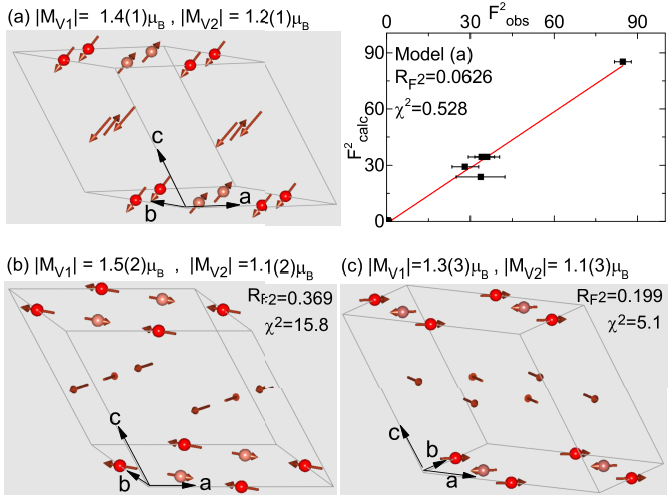}
\caption {(a)-(c): Three magnetic structure models showing ferrimagnetism with different easy axis. Model (a) is collinear suggesting weak magnetic anisotropy while Models (b) and (c) are noncollinear indicating strong magnetic anisotropy. Model (a) is the magnetic structure of $\beta$-V$_2$PO$_5$. }\label{fig2}
\end{figure}

The magnetic order onsets at 128 K while the charge order continues to develop below the magnetic transition. Since no observed sharp change can be determined by the structure refinement at 4 K (see Table I), no further structural phase transition below 128 K is discernable. The magnetic propagation vector is $k$=0, which means that the magnetic scattering signal appears on top of the nuclear Bragg peaks. To determine the magnetic structure more precisely, the magnetic signals were extracted by subtracting the data measured just above 128 K from that at 4 K. During the procedure, to select peaks which are insensitive to the thermal displacements and charge ordering, we compared the data measured at 300 K and 450 K and only selected a peak if the change of its intensity is much smaller than the extracted magnetic intensity. The selected reflections are listed in Table SI \cite{supp}. Since (1 1 0)$_T$, (1 1 4)$_T$, (1 0 1)$_T$, (0 0 2)$_T$ and (0 0 4)$_T$ peaks were measured with a long counting time and also tracked upon warming, they are highly reliable as indicated by their small error bars shown in the Table SI. We then performed the representational analysis that determines the symmetry-allowed magnetic structures for a second-order magnetic transition. It yielded two magnetic symmetries, $C2/c$ and $C2'/c'$. Only the $C2'/c'$ can fit our data (see Table SI). The obtained magnetic structure is ferrimagnetic. Spins on all V$^{2+}$ (V1 sites) atoms are parallel and so do the spins on all V$^{3+}$ (V2 sites) atoms while these two spin sublattices are antiparallel to each other. Since it is unlikely for the V$^{2+}$ ($d^3$) to be in a low spin state due to the longer V-O bond length on V1 site and thus weaker crystal electric field, the moment $M_{V1}>M_{V2}$. Figure 4 shows the ferrimagnetic structure with three possible easy axis assignments where $M_{V1}>M_{V2}$. The calculated peak intensity and goodness of fit are summarized in Table SI. The model in the left panel of Fig. 4(a) is our pick for the $\beta$-V$_2$PO$_5$ which gives the best fit of the data as shown in the right panel of Fig. 4(a). The $M_{V1}= 1.4(1) \mu_B$ and $M_{V2}= 1.2(1) \mu_B$. The easy axis is in $ac_M$ plane and 47(9) degrees away from $a_M$ towards $c_M$. The magnetic structure is collinear, suggesting weak magnetic anisotropy. This is indeed consistent with the anisotropic M(H) measurements shown in Fig. 1(c). The other two models shown in Fig. 4(b) and (c) are with the easy axis along the chain direction (Fig. 4(b)) or perpendicular to the chain direction on the $ab$ plane (Fig. 4(c)). These two magnetic models are non-collinear with strong magnetic anisotropy. Since the goodness of fit for these two latter models are poor, they are not the right magnetic structure for $\beta$-V$_2$PO$_5$.

\section{Discussion}

\begin{figure}
\includegraphics[width=3.5in]{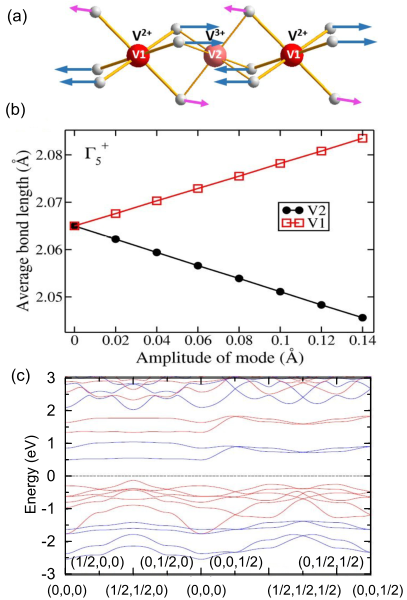}
\caption {(a) A sketch of the Oxygen anion displacements due to the $\Gamma_5^+$ mode which is responsible for charge ordering. (b) The average $V-O$ bond length as a function of the $\Gamma_5^+$ normal mode amplitude. (c) DFT band structure in the ferrimagnetic state. Majority and minority spin bands are shown in red and blue respectively.
}\label{fig_theory}
\end{figure}

It is of particular interest to ask if the charge order is solely responsible for the reduction in the crystal symmetry, or rather if it is a secondary order parameter to some other electronic phase transition. In order to preclude this possibility and elucidate the nature of the charge ordering transition at 610 K, we performed a group theoretical analysis of the lattice distortion using the Isotropy Software Suite.\cite{isotropy}
The distortion from the high temperature tetragonal structure ($I4_1/amd$) to the low temperature monoclinic structure ($C2/c$) can be caused by two normal modes of lattice vibration, $\Gamma_5^+$ and $\Gamma_4^+$.
$\Gamma_5^+$ reduces the symmetry from $I4_1/amd$ to $C2/c$, and is the only candidate for the primary structural order parameter, whereas $\Gamma_4^+$ by itself reduces the symmetry to $Fddd$. $C2c$ is a subgroup of $Fddd$, and as a result, $\Gamma_4^+$ is most likely a secondary order parameter that is not important in the energetics of the phase transition. At the same time, the charge order itself, which is a differentiation of the neighboring V ions in the same chain, transforms as the $\Gamma_5^+$ irreducible representation does for the high symmetry structure. Figure \ref{fig_theory}(a) shows the displacement of the oxygen atoms in the VO$_6$ face-sharing chain due to the $\Gamma_5^+$ normal mode. This mode breaks the symmetry between the V ions that are symmetry equivalent at $I4_1/amd$, and decreases the V-O bond length for $V_2$ while increasing it for $V_1$ (Fig. \ref{fig_theory}(b)) as expected in a charge ordering transition. We therefore conclude that to reduce the symmetry to the monoclinic phase, the charge order, by itself, is sufficient and no other magnetic or electronic mechanisms are necessary. This is consistent with the fact that the $T_{CO}$ is almost 5 times of the $T_{mag}$. We also note that $\Gamma_5^+$ is a Raman active mode, and as a result, signature of the charge ordering transition should be visible in the Raman spectrum of $\beta$-V$_2$PO$_5$.

We performed the first principles calculation using DFT+U with U = 4 eV to correct for the underestimation of the on-site coulomb interaction on the V ion \cite{DFTU}. Our DFT calculations predict magnetic moments of 2.6 $\mu_B$ and 1.8 $\mu_B$ respectively inside the $V_1$ and $V_2$ spheres, but the band structure (Fig. \ref{fig_theory}(c)) shows no partially filled bands. This signals strong hybridization between the $V$ and the $O$ ions. DFT gives a magnetic moment 0.5 $\mu_B$ per $V$ including the interstitials. This is a strong overestimation compared to the experimental value. The reason of this is likely the DFT+U's tendency to overestimate the ordered moments when there are dynamical fluctuations present, and it is possible that a more advanced first principles method (such as the Dynamical Mean Field Theory) can reproduce the experimentally observed value of the local moments.

\begin{table}
\caption{Energies of different magnetic configurations from first principle.  }
\begin{tabular}{c|c|c}
\hline
\hline
Intra-chain  &Inter-chain  & Energy (meV/f.u.)\\
\hline
Ferrimagnetic	& Ferromagnetic	& 0\\
Ferrimagnetic	& Antiferromagnetic	& 9\\
Ferromagnetic	& Ferromagnetic	& 60\\
Ferromagnetic	& Antiferromagnetic	& 71 \\

\hline
\hline

\end{tabular}

\label{table_theory}
\end{table}

To understand the magnetic order, we calculated the energies of phases with different magnetic orders from DFT, as listed in Table \ref{table_theory}. The lowest energy phase is predicted to have ferrimagnetic intra-chain order, where the moments of the neighboring $V_1$ and $V_2$ ions on the same chain are aligned anti-parallel, and ferromagnetic inter-chain order, so that, for example, the magnetic moments of all $V_1$ ions are parallel. This observation is in line with the experimental observation. The energy cost of having a magnetic phase where the different chains have antiparallel moments is bout 9-10 meV per formula unit, whereas the energy cost of having spins on the same chain parallel is ~60 meV per formula unit. This suggests that the intra-chain interactions between the $V_1$ and $V_2$ ions in the face-sharing VO$_6$ chains are the dominant term in the magnetic hamiltonian.

To understand the crossover behavior in H/M shown in in Fig. 1(b), we calculated the energetics of different magnetic phases in the high temperature tetragonal structure ($I4_1/amd$) with the same parameters (not shown), and found that similar couplings apply to magnetic moments in that structure too. This explains the cross-over: Above the charge ordering temperature, the magnetic moments on all V ions are equal, and its Curie-Weiss behaviour is that of an antiferromagnet, with a negative Curie temperature. However, charge ordering makes the moments unequal, and as a result, below 610 K the Curie-Weiss behaviour is that of an ferrimagnet, which has a positive Curie temperature like a ferromagnet.

To address if there is non-trivial topology in this compound, the DFT band structure in Fig. \ref{fig_theory}(c) is calculated in the ferrimagnetic ground state. Unlike the DFT band structure calculated in the ferromagnetic phase and the tetragonal structure without U \cite{v2o5p}, there are no band crossing at the Fermi level. And more importantly, the top of the valence and the bottom of the conduction bands have opposite spin directions. This observation precludes any possibility of topological phases in $\beta$-V$_2$PO$_5$.

\section{Conclusion}

In conclusion, we have grown and characterized $\beta$-V$_2$PO$_5$ single crystals. A tetragonal to monoclinic structural phase transition at 610 K and a paramagnetic to ferrimagnetic phase transition are revealed by transport, magnetic, specific heat, single crystal neutron diffraction and NMR measurements. Below 610 K, the single V site at the high temperature tetragonal phase distorts into two alternating V sites, leading to the increase of V-O bond length of one V site but the decrease of the other V site. Our first principles calculation shows that this distortion was caused by the $\Gamma_5^+$ normal mode of lattice vibration. Accompanied with the distortion, charge ordering of V$^{2+}$ and V$^{3+}$ is undoubtedly suggested by the Bond-valence analysis. Below 128 K, the spins order parallel on each sublattice of V sites while the spins on one sublattice order antiparallel to the other. With the easy axis being in the monoclinic $ac$ plane and 47(9)$^\circ$ away from the monoclinic $a$ towards $c$ axis., this gives a collinear ferrimagnetic structure with the moment to be 1.4(1)$\mu_B$ on V$^{2+}$ site and 1.2(1)$\mu_B$ on V$^{3+}$ site, suggesting weak magnetic anisotropy and dominant role of the intra-chain V-V interaction in magnetism. No non-trivial topology is suggested by our first principles calculation.

\begin{acknowledgments}
Work at UCLA (JX, CWH, RC, NN) was supported by NSF DMREF program under the award NSF DMREF project DMREF-1629457.
Work at ORNL HFIR was sponsored by the Scientific User Facilities Division, Office of Science, Basic Energy Sciences, U.S. Department of Energy.
Work at UMN was supported by NSF DMREF program under the award NSF DMREF project DMREF-1629260.
Work at UCLA (HHW, YKL, SB) was supported by NSF DMR-1410343 and DMR-1709304. YKL would also like to thank the support from LANL LDRD program.

\end{acknowledgments}

\textit {Note}: During the preparation of this manuscript, we noticed a work on pollycrystalline $\beta$-V$_2$OPO$_4$ was just accepted but Journal of the american chemistry society (http://pubs.acs.org/doi/abs/10.1021/jacs.7b09441), which also revealed the charge ordering and similar ferrimagnetism with different easy axis in this compound.


\begin{thebibliography}{99}
\bibitem{COTO} J. Attfield. Charge ordering in transition metal oxides. \emph{Solid state sciences} \textbf{8}, 861 (2006).
\bibitem{MnO1} E. Wollan, W. Koehler. Neutron Diffraction Study of the Magnetic Properties of the Series of Perovskite-Type Compounds [(1-x)La,xCa]MnO$_3$. \emph{Phys. Rev.} \textbf{100}, 545 (1955).
\bibitem{MnO2}  J. Goodenough. Theory of the Role of Covalence in the Perovskite-Type Manganites [La,M(II)]MnO$_3$. \emph{Phys. Rev.} \textbf{100}, 564 (1955).
\bibitem{vosc} T. Yamauchi, M. Isobe, and Y. Ueda. Charge order and superconductivity in vanadium oxides. \emph{Solid State Sciences} \textbf{7}, 874 (2005).

\bibitem{vpo4} R. Glaum, M. Reehuis, N. St$\ddot{u}\beta$er, U. Kaiser, and F. Reinauer. Neutron Diffraction Study of the Nuclear and Magnetic Structure of the CrVO$_4$ Type Phosphates TiPO$_4$and VPO$_4$. \emph{J. Solid State Chem.} \textbf{126}, 15 (1996).
\bibitem{VO(PO3)2nmr} J. Kikuchi, N. Kurata, K. Motoya, T. Yamauchi, and Y. Ueda. Spin Diffusion in the S=1/2 Quasi One-Dimensional Antiferromagnet $\alpha$-VO(PO$_3$)$_2$ via 31P NMR. \emph{J. Phys. Soc. Jpn.} \textbf{70}, 2765 (2001).
\bibitem{v2vop2o72} J. Johnson, D. Johnston, H. King Jr., T. Halbert, J. Brody, and D. Goshorn. Structure and magnetic properties of V$_2$(VO)(P$_2$O$_7$)$_2$. A mixed-valence vanadium (III, III, IV) pyrophosphate. \emph{Inorg. Chem.} \textbf{27}, 1646 (1988).
\bibitem{(VO)2P2O7 theory1} D. Johnston, J. Johnson, D. Goshorn, and A. Jacobson. Magnetic susceptibility of (VO)$_2$P$_2$O$_7$: A one-dimensional spin-1/2 Heisenberg antiferromagnet with a ladder spin configuration and a singlet ground state. \emph{Phys. Rev. B} \textbf{35}, 219 (1987).
\bibitem{(VO)2P2O7 theory2} E. Dagotto, J. Riera, and D. Scalapino. Superconductivity in ladders and coupled planes. \emph{Phys. Rev. B} \textbf{45} ,5744 (1992).
\bibitem{(VO)2P2O7 neotron} A. W. Garrett, S. E. Nagler, D. A. Tennant, B. C. Sales, and T. Barnes. Magnetic excitations in the S=1/2 alternating chain compound (VO)$_2$P$_2$O$_7$. \emph{Phys. Rev. Lett.} \textbf{79}, 745 (1997).
\bibitem{(VO)2P2O7nmr} J. Kikuchi, K. Motoya, T. Yamauchi, and Y. Ueda. Coexistence of double alternating antiferromagnetic chains in (VO)$_2$P$_2$O$_7$: NMR study. \emph{Phys. Rev. B} \textbf{60}, 6731 (1999).
\bibitem{(VO)2P2O7nmr2} T. Yamauchi, Y. Narumi, J. Kikuchi, Y. Ueda, K. Tatani, T. C. Kobayashi, K. Kindo, and K. Motoya, \emph{Phys. Rev. Lett.} \textbf{83}, 3729 (1999).
\bibitem{v2o5p structure1} R. Glaum, and R. Gruehn. Synthese, Kristallstruktur und magnetisches Verhalten von V2PO5. \emph{Z. Kristallogr} \textbf{186}, 91 (1989).
    \bibitem{johnston} D. C. Johnston, Magnetic Susceptibility of Collinear and Noncollinear Heisenberg Antiferromagnets, Phy. Rev. Lett., 109, 077201 (2012)

\bibitem{v2o5p} Y. Jin, R. Wang, Z. Chen, J. Zhao, Y. Zhao, and H. Xu. Ferromagnetic Weyl semimetal phase in a tetragonal structure. \emph{Phys. Rev. B} \textbf{96}, 201102 (2017).

\bibitem{silver} D. Smith, and F. Fickett. Low-temperature properties of silver. \emph{J. Res. Natl. Inst. Stand. Technol.} \textbf{100}, 119 (1995).
\bibitem{hb3a} B. Chakoumakos, H. Cao, F. Ye, A. Stoica, M. Popovici, M. Sundaram, W. Zhou, J. Hicks, G. Lynn and R. Riedel. Four-circle single-crystal neutron diffractometer at the High Flux Isotope Reactor. \emph{J. Appl. Crystallogr.} \textbf{44}, 655 (2011).
\bibitem{sarah} A. Wills. A new protocol for the determination of magnetic structures using simulated annealing and representational analysis (SARA\emph{h}). \emph{Phys. B: Condens. Matt.} {\bf 276}, 680 (2000), program available from www.ccp14.ac.uk.
\bibitem{fullprof} J. Rodriguez-Carvajal. Recent advances in magnetic structure determination by neutron powder diffraction. \emph{Phys. B: Condens. Matt.} \textbf{192}, 55 (1993).
\bibitem{VASP1}G. Kresse and J. Furthm{\"u}ller. Efficient iterative schemes for ab initio total-energy calculations using a plane-wave basis set. \emph{Phys. Rev. B} \textbf{54}, 11169 (1996).
\bibitem{VASP2} G. Kresse, and J. Hafner. Ab initio molecular dynamics for open-shell transition metals. \emph{Phys. Rev. B} \textbf{48}, 13115 (1993).
\bibitem{PBEsol} J. Perdew, A. Ruzsinszky, G. Csonka, O. Vydrov, G. Scuseria, L. Constantin, X. Zhou, and K. Burke. Restoring the Density-Gradient Expansion for Exchange in Solids and Surfaces. \emph{Phys. Rev. Lett.} \textbf{100}, 136406 (2008).
   \bibitem{supp} See supplimentary material.
\bibitem{arrot1} A. Arrott, and J. Noakes. Approximate equation of state for nickel near its critical temperature. \emph{Phys. Rev. Lett.} \textbf{19}1,786 (1967).
\bibitem{arrot2} M. Halder, S. M. Yusuf, M. D. Mukadam, and K. Shashikala, Phys. Rev. B 81, 174402 (2010)
\bibitem{entropy dimension1} L. Jongh, and A. Miedema. Experiments on simple magnetic model systems. \emph{Advances in Physics} \textbf{23},1 (1974).
\bibitem{entropy dimension2} S. Nagata, P. Keesom, and S. Faile. Susceptibilities of the vanadium Magn$\acute{e}$li phases V$_n$O$_{2n-1}$ at low temperature. \emph{Phys. Rev. B} \textbf{20}, 2886 (1979).
\bibitem{VPOglass} M. Khan, R. Harani, M. Ahmed, and C. Hogarth. A comparative study of the effects of rare-earth oxides on the physical, optical, electrical and structural properties of vanadium phosphate glasses. \emph{J. Materials Science} \textbf{20}, 2207 (1985).
\bibitem{NMR1} M. Sananes, and A. Tuel. Study by 31 P NMR spin echo mapping of vanadium phosphorus oxide catalysts. \emph{Solid state nuclear magnetic resonance} \textbf{6}, 157 (1996).
\bibitem{NMR2} M. Sananes, A. Tuel, and J. C. Volta. A study by 31P NMR spin-echo mapping of VPO catalysts: I. characterization of the reference phases. \emph{J. Catalys.} \textbf{145}, 251 (1994).
\bibitem{NMR3} J. Li, M. Lashier, G. Schrader, B. Gerstein. Oxidation states of vanadium in V-P-O oxidation catalysts 31P NMR by spin-echo mapping. \emph{Appl. Catalys.} \textbf{73}, 83 (1991).
\bibitem{MORIYA} T. Moriya. Nuclear magnetic relaxation in antiferromagnetics. \emph{Prog. Theor. Phys.} \textbf{16}, 23 (1956).
\bibitem{bondvalence_parameters} N. Brese, and M. O'keeffe. Bond-valence parameters for solids. \emph{Acta Crystallographica Section B: Structural Science} \textbf{47}, 192 (1991).
\bibitem{bondvalence_parameters1} I. David Brown, Bond valence parameters, https://www.iucr.org/resources/data/datasets/bond-valence-parameters
\bibitem{isotropy} H. Stokes, D. Hatch and B. Campbell. Isotropy. Retrieved from stokes. byu. edu/isotropy. html(2007).
\bibitem{DFTU}A. Liechtenstein, V. Anisimov and J. Zaanen. Density-functional theory and strong interactions: Orbital ordering in Mott-Hubbard insulators. \emph{Phys. Rev. B}, \textbf{52}, R5467 (1995).




\end{thebibliography}
\end{document}